\documentclass[10pt,twocolumn,letterpaper]{article}

\usepackage{placeins}   

\usepackage{booktabs}
\usepackage{siunitx}
\usepackage{cvpr}

\usepackage{amsmath,amssymb,mathtools}
\usepackage{bm}
\usepackage{xspace}
\usepackage{xcolor}
\usepackage[table]{xcolor}

\newcommand{\name}{C$^{2}$T\xspace}

\usepackage{algorithm}
\usepackage{algpseudocode}

\usepackage{times,epsfig,graphicx}
\usepackage{amsmath,amssymb,mathtools,bm}
\usepackage{algorithm}
\usepackage{algpseudocode}



\definecolor{cvprblue}{rgb}{0.21,0.49,0.74}
\usepackage[pagebackref,breaklinks,colorlinks,allcolors=cvprblue]{hyperref}

\def\paperID{43337}
\def\confName{CVPR}
\def\confYear{2026}

\title{C$^2$T: Captioning-Structure and LLM-Aligned Common-Sense Reward Learning for Traffic--Vehicle Coordination}

\author{
Yuyang Chen$^{1}$, Kaiyan Zhao$^{2}$, Yiming Wang$^{4}$, Ming Yang$^{3}$, Bin Rao$^{1}$, Zhenning Li$^{1*}$\\[0.5em]
$^{1}$The State Key Laboratory of Internet of Things for Smart City, University of Macau\\
$^{2}$Wuhan University \quad
$^{3}$Hong Kong Polytechnic University\\
$^{4}$Computer and Information Science, University of Macau\\
{\tt\small zhenningli@um.edu.mo}
}

\begin{document}
\maketitle

\begin{abstract}
State-of-the-art (SOTA) urban traffic control increasingly employs Multi-Agent Reinforcement Learning (MARL) to coordinate Traffic Light Controllers (TLCs) and Connected Autonomous Vehicles (CAVs). However, the performance of these systems is fundamentally capped by their hand-crafted, myopic rewards (e.g., intersection pressure), which fail to capture high-level, human-centric goals like safety, flow stability, and comfort. To overcome this limitation, we introduce C2T, a novel framework that learns a common-sense coordination model from traffic–vehicle dynamics. C2T distills “common-sense” knowledge from a Large Language Model (LLM) into a learned intrinsic reward function. This new reward is then used to guide the coordination policy of a cooperative multi-intersection TLC MARL system on CityFlow-based multi-intersection benchmarks. Our framework significantly outperforms strong MARL baselines in traffic efficiency, safety, and an energy-related proxy. We further highlight C2T’s flexibility in principle, allowing distinct “efficiency-focused” versus “safety-focused” policies by modifying the LLM prompt.
\end{abstract}
\section{Introduction}
\label{sec:intro}

Urban traffic control has long been a central topic in intelligent transportation. Reinforcement-learning (RL)–based traffic light controllers typically optimize local efficiency by minimizing queue length, intersection pressure, or average delay~\cite{wei2018intellilight,wei2019presslight}. While such handcrafted objectives can deliver short-term gains, they often overlook network-level dynamics. When each intersection pursues its own local optimum, coordination across the city breaks down, producing congestion waves, unstable flows, and frequent braking. In practice, controllers that aggressively clear volume may underweight smoothness, safety, and comfort, yielding traffic that looks efficient on paper but feels brittle and unpredictable in deployment.

Coordinated control aims to overcome these limitations by letting signals and vehicles shape one another’s behavior: signals create openings and platoons, while vehicles adjust approach speeds and lane usage to match those plans. CoTV-style systems~\cite{guo2023cotv} and related work on coordinated signal control~\cite{wei2019presslight} show that such coordination can improve travel time and energy in simulation, even when vehicles themselves are not explicitly modeled as learning agents. Yet the underlying reward recipe is usually unchanged: signals are scored by pressure or queue imbalance, and vehicles by speed tracking and acceleration penalties. These proxies favor aggressive clearing but say little about stability, safety, or comfort, leading to oscillatory phase changes, abrupt braking, and unsafe spacing even when average efficiency improves. What is missing is a notion of \emph{traffic quality} that reflects human judgment and anticipates longer-horizon effects (e.g., platoon formation), and that can be learned offline while remaining compatible with standard multi-intersection RL training.

We take traffic quality itself as supervision and gather it before training. Evidence from other domains shows that large language models (LLMs) and related systems can make consistent pairwise judgments when asked to compare well-structured state descriptions~\cite{christiano2017deep,ouyang2022instructgpt}. Building on this observation, we propose \textbf{C$^2$T}, a framework that renders each observation into a deterministic, unit-aware caption; samples caption pairs across time and approaches; and queries an LLM under a fixed prompt to decide which state better reflects good traffic. We retain only decisive labels and fit a lightweight preference scorer that outputs an intrinsic reward, with optional heads emphasizing safety, efficiency, or energy. All supervision is produced offline and cached. At training time, the intrinsic signal is added to a standard multi-intersection PPO pipeline as an additional reward stream—no simulator changes and no online LLM calls. In this paper, we instantiate \textbf{C$^2$T} on CityFlow-based benchmarks~\cite{zhang2019cityflow} where each intersection-level traffic-light controller (TLC) is a learning agent and vehicles follow the built-in microscopic dynamics, but the same design extends naturally to settings with explicitly modeled connected autonomous vehicles (CAVs).

A key design choice is to mix the intrinsic reward \emph{asymmetrically}: TLCs receive a blend of environment reward and the learned score, whereas vehicles remain environment driven in our instantiation. This places the human-aligned guidance where phase selection most strongly shapes platoons and network rhythm, and avoids extra non-stationarity that would arise if both sides were driven by a changing intrinsic signal. Safety and stability are enforced by a lightweight risk mask that suppresses the intrinsic signal under low time-to-collision (TTC) percentiles, clusters of harsh deceleration, or red-light risk flags parsed from the caption. For optimization, each reward stream is normalized and softly clipped before advantage estimation, and the mixing weight is scheduled so the policy first satisfies environment constraints and then gradually absorbs common-sense preferences. Because the scorer is trained offline, switching prompts or enabling safety/efficiency/energy heads only changes the cached scores, leaving the RL loop unchanged.

Finally, we evaluate \textbf{C$^2$T} on CityFlow-based benchmarks following the LLMLight protocol. We consider three real-city networks (Jinan, Hangzhou, and New York) and two stress-test scenarios (extreme demand surges and 24-hour diurnal cycles), and we reuse the same training pipeline and random seed protocol for fair comparison. Across demand levels and seeds, \textbf{C$^2$T} reduces average travel time and increases throughput while improving safety, as indicated by higher TTC percentiles and fewer harsh-braking events; an energy-related proxy also improves. Ablations show that the intrinsic signal, the safety mask, and the combination of per-stream normalization with a simple schedule are all necessary for stable gains. Captions alone already help, and adding matched numeric slots yields further (smaller) improvements—underscoring the value of structured descriptions for learning traffic quality offline and integrating it into existing multi-intersection RL training without altering the simulator or incurring online LLM calls.

In summary, this work delivers:
\begin{itemize}\itemsep 0pt \parskip 0pt \parsep 0pt
    \item \textbf{Structured captioning for traffic RL:} a deterministic, unit-aware schema that makes state quality comparable by LLMs and reproducible across runs.
    \item \textbf{Offline preference learning for an intrinsic reward:} a lightweight scorer trained from decisive LLM preferences on caption pairs, with optional heads for safety, efficiency, and energy, and no online LLM calls.
    \item \textbf{Safe, asymmetric integration in multi-intersection RL:} intrinsic reward mixed only into the TLC objective, combined with a risk mask, per-stream normalization, soft clipping, and a simple schedule—fully compatible with standard PPO.
    \item \textbf{Evidence across benchmarks:} gains on CityFlow-based Jinan, Hangzhou, and New York networks, plus stress scenarios and ablations that clarify the role of the intrinsic signal, safety mask, and optimization choices.
\end{itemize}

\section{Related Work}
\label{sec:related}

\subsection{Multi-Agent Reinforcement Learning for Traffic Signal Control}

RL-based traffic signal control started from single-intersection agents with handcrafted efficiency rewards such as delay or queue length, e.g., IntelliLight and the max-pressure-based PressLight~\cite{wei2018intellilight,wei2019presslight}. Later work focuses on coordination and scalability: CoLight propagates information via graph attention, FRAP models phase competition, and MPLight scales pressure-based control to city networks~\cite{wei2019colight,zheng2019frap,chen2020mplight}, with further refinements through attention architectures, meta-learning, and improved pressure features in AttendLight, MetaLight, Efficient-CoLight, and Advanced-CoLight~\cite{orooglooyjadid2020attendlight,zang2020metalight,wu2021efficientcolight,zhang2022advancedcolight}. 
CoTV additionally couples TLC and CAV control under a composite reward over travel time, fuel, and emissions~\cite{guo2023cotv}, but all these methods ultimately optimize fixed metric-based scalar rewards, whereas \name{} learns a commonsense intrinsic reward from LLM preferences over structured TLC--CAV descriptions. More broadly, recent RL work has improved learning via better experience replay~\cite{zhao2025eder}, latent exploration~\cite{wang2025bile,wang2026lspe,wang2026exploretolearn}, and goal-conditioned generalization~\cite{wang2026dsap}. These methods, however, focus on generic sample efficiency, exploration, or transfer rather than human-aligned traffic-quality reward design.
\subsection{Reward Modeling and Preference Learning in Reinforcement Learning}

Handcrafted rewards are fragile proxies for designer intent and often induce reward hacking in complex domains such as urban traffic. IRL methods like GAIL and AIRL infer rewards from expert demonstrations~\cite{ho2016gail,fu2018airl}, while preference-based RL learns a reward model from pairwise human comparisons~\cite{christiano2017deep}, but both are hard to scale to large multi-agent traffic systems due to data demands, ambiguity, and computational cost. \name{} instead adopts an AI-assisted preference-learning view: it queries an LLM on deterministic, unit-aware captions and distills its judgments into an intrinsic reward tailored to TLC--CAV coordination.

\subsection{Large Language Models for Decision Making and Reward Alignment}

Recent work uses LLMs as high-level decision makers by mapping states to text and decoding actions, e.g., LLMLight for traffic signals and holistic vision--language driving with counterfactual reasoning in OmniDrive~\cite{lai2023llmlight,simavqadrive2025}, but directly placing LLMs in the control loop raises latency, reliability, and scalability issues in multi-agent settings. A complementary line treats LLMs as judges: DriveLM evaluates driving decisions via graph-based QA~\cite{simadrivelm2024}, and RLHF/RLAIF-style approaches train policies against reward models learned from human or AI feedback~\cite{ouyang2022instructgpt,xu2025llmjudge}, with Poutine-style systems showing that vision--language models plus light preference tuning can yield strong driving performance~\cite{xu2025llmjudge}. Existing LLM-as-judge work, however, typically operates on free-form descriptions or QA pairs for single agents, whereas \name{} uses deterministic schema-based captions to obtain stable pairwise preferences that can be distilled into a reusable intrinsic reward for MARL.

\subsection{Structured Scene Representation and Driving Captions}

Graph-based and object-centric representations factor traffic scenes into entities and relations, and are widely used in RL for phases, lanes, and movements, as in FRAP and CoLight~\cite{zheng2019frap,wei2019colight}, but these formats are designed for neural nets rather than LLMs. Vision-based driving captioning and VQA datasets, such as nuScenes caption/QA benchmarks and DriveLM, provide natural language descriptions of driving scenes, and OmniDrive extends them to counterfactual reasoning~\cite{inoue2024nuscenescaption,simadrivelm2024,simavqadrive2025}; yet free-form captions are often stylistic and under-specified, lacking precise quantitative details needed for consistent reward evaluation. \name{} bridges this gap by constructing deterministic, unit-aware, schema-constrained traffic captions that enumerate key variables (e.g., queues, delays, TTC, violations) with explicit semantics, making them both faithful to the underlying state and well-suited for stable LLM-based preference learning.

\section{Preliminaries}
\label{sec:preliminaries}

We briefly review the reinforcement learning formulation for traffic signal control, from the standard single-agent MDP view to a cooperative multi-agent setting with traffic lights (TLCs) and connected autonomous vehicles (CAVs). This section only fixes notation and highlights the limitations of handcrafted rewards that motivate C$^2$T.

\subsection{Traffic Signal Control as an MDP}
\label{sec:tlc_mdp}

We model a single intersection as a Markov Decision Process
$\mathcal{M}=(\mathcal{S},\mathcal{A},\mathcal{P}, r_{\text{ext}}, \gamma)$.
At time step $t$, the TLC observes a local traffic state
$s_t \in \mathcal{S}$ containing the current phase, its elapsed time, and lane-level statistics such as queue length and flow.
It then chooses an action $a_t \in \mathcal{A}$ (e.g., \texttt{keep} or \texttt{switch} phase), and the microscopic simulator applies
$\mathcal{P}(s_{t+1}\mid s_t,a_t)$.

The external reward $r_{\text{ext}}(s_t,a_t)$ is a handcrafted congestion proxy.
A common choice is negative intersection pressure:
\begin{equation}
r_{\text{ext}}(s_t)
= - \sum_{i=1}^{K} \big( n^{in}_i - n^{out}_i \big),
\label{eq:pressure_reward}
\end{equation}
where $n^{in}_i$ and $n^{out}_i$ are the inflow and outflow vehicle counts on approach $i$.
The TLC learns a policy $\pi(a_t\mid s_t)$ that maximizes the discounted return
$\mathbb{E}_\pi[\sum_{t=0}^{\infty} \gamma^t r_{\text{ext}}(s_t,a_t)]$.

This formulation has been widely used in RL-based traffic signal control, but it implicitly treats vehicles as passive entities and only optimizes short-term efficiency metrics defined by $r_{\text{ext}}$.

\subsection{Multi-Agent Traffic Control and Its Limitations}
\label{sec:marltlc}

Modern intelligent transportation systems increasingly coordinate both infrastructure and vehicles. We therefore consider a multi-agent setting with TLC agents $M=\{1,\dots,|M|\}$ and (potential) CAV agents $N=\{1,\dots,|N|\}$.
At time $t$, each agent $i\in M\cup N$ receives a local observation $o_t^i$, selects an action $a_t^i$, and the shared traffic state evolves as
\[
s_{t+1} = f\!\left(s_t, \{a_t^i\}_{i \in M \cup N}\right).
\]

TLC agents aim to reduce local congestion via external rewards such as
\[
r^{\text{TL}}_{\text{ext}}(s_t)
= - \sum_{i=1}^{K} \big( n^{in}_i - n^{out}_i \big)
  - \lambda \cdot \text{delay}_t,
\]
while CAV agents, when explicitly controlled, are typically encouraged to track a desired speed with smooth accelerations, e.g.,
\[
r^{\text{CAV}}_{\text{ext}}(s_t)
= - \big|v_t - v^* \big|
  - \beta \, \|a_t\|.
\]

The cooperative objective is to optimize all agents jointly:
\begin{equation}
\max_{\pi_M,\, \pi_N} \;
\mathbb{E} \Bigg[
\sum_{t=0}^{T}
\sum_{i \in M \cup N}
\gamma^t \, r_t^i
\Bigg],
\label{eq:jointobj}
\end{equation}
where $\pi_M$ and $\pi_N$ are the (shared) policies for TLCs and CAVs.

However, both the single-agent and multi-agent formulations above are still driven by simple, handcrafted rewards $r^{\text{TL}}_{\text{ext}}$ and $r^{\text{CAV}}_{\text{ext}}$ that focus on local pressure, delay, or speed tracking. They provide limited guidance about broader notions of safety, comfort, and flow stability, and may therefore produce policies that are numerically optimal yet misaligned with human common sense.

This motivates our central goal: to learn a \emph{human-aligned} intrinsic reward $r_\phi$ from preferences expressed by large language models, and to integrate it with the external rewards in multi-agent traffic control. The next section introduces the C$^2$T framework that realizes this idea.

\section{Method}
\label{sec:method}

\subsection{Overview}
\label{sec:overview}
We propose \textbf{C$^2$T} (\textit{Captioning-Structure and LLM-Aligned Common-Sense Reward Learning}).
C$^2$T has three stages:
(\textbf{Stage~1}) render simulator observations into structured captions and build preference pairs;
(\textbf{Stage~2}) learn a common-sense intrinsic reward $r_\phi$ from LLM preferences;
(\textbf{Stage~3}) inject $r_\phi$ into the RL loop via asymmetric reward mixing with explicit safety masks.
We instantiate C$^2$T on CityFlow-based multi-intersection benchmarks where each traffic-light controller (TLC) is a learning agent while vehicles follow the built-in microscopic dynamics.
In this work we focus on TLC-only learning and discuss extensions to explicitly modeled connected autonomous vehicles (CAVs) in the appendix.

\vspace{2pt}
\begin{figure*}[t]
  \centering
  \includegraphics[width=\linewidth]{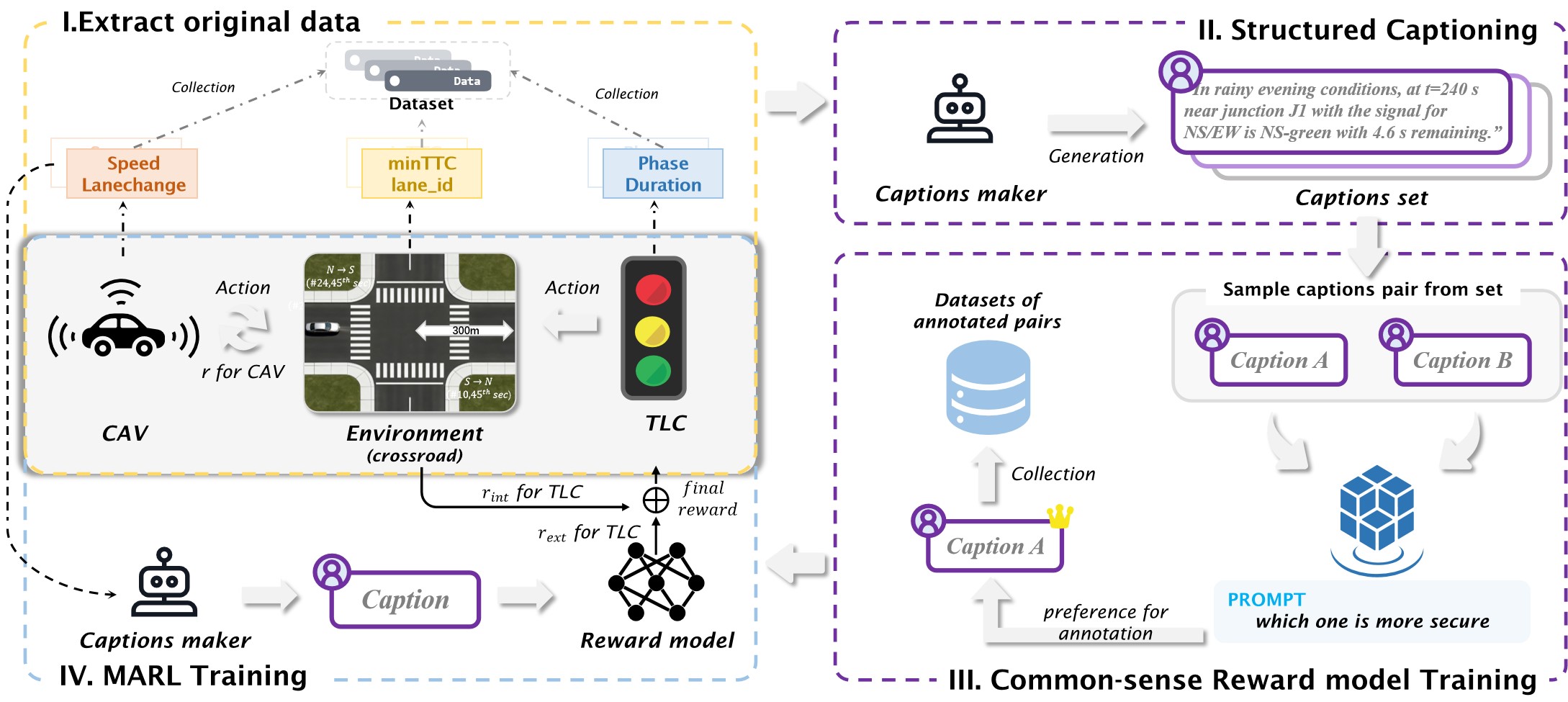}
  \caption{C$^2$T pipeline. Stage~1 converts raw observations into schema-constrained captions and samples high-contrast/safety-contrast pairs. Stage~2 queries an LLM to label pairwise preferences and trains a scalar scorer $r_\phi$ with a Bradley--Terry loss (with frequency reweighting and score centering). Stage~3 freezes $r_\phi$ and injects it as an intrinsic component for TLC steps under a safety mask, mixing with external rewards; all streams are normalized before GAE+PPO. In our CityFlow instantiation, only TLCs are learning agents and vehicles remain part of the environment dynamics.}
  \label{fig:c2t_pipeline}
\end{figure*}
\vspace{-6pt}

Figure~\ref{fig:c2t_pipeline} summarizes the end-to-end pipeline. We next detail each stage.

\subsection{Stage 1: Structured Captioning and Preference Pairs}
\label{sec:stage1_caption}

Let $o\in\mathcal{O}$ be an observation at time $t$ for a given intersection.
We define a deterministic captioner
\begin{equation}
\label{eq:captioner}
c \;=\; \mathcal{C}(o;\,\Xi),
\end{equation}
where $\Xi$ is a fixed schema (fields and units). The schema concatenates structured slots (no free-form prose):
\begin{equation}
\label{eq:caption_schema}
\begin{aligned}
c = \big[\, &
\underbrace{\text{phase},\,\text{elapsed}}_{\text{signal}},\;
\underbrace{\{q_i,\,p_i\}_{i=1}^{K}}_{\text{queues/pressure}},\;
\underbrace{\bar d,\,\bar \tau}_{\text{delay/throughput}}, \\
& \underbrace{\mathrm{TTC}_{\mathrm{p10}},\,\mathrm{TTC}_{\mathrm{p50}},\,h_{\text{brake}},\,\rho_{\text{red}}}_{\text{safety}},\;
\underbrace{v_{\text{near}},\,a_{\text{near}},\,d_{\text{stop}}}_{\text{nearest vehicle}}
\,\big].
\end{aligned}
\end{equation}
Here $q_i$ and $p_i$ denote queue and pressure on approach $i$, $\bar d$ average delay, $\bar \tau$ throughput,
$\mathrm{TTC}_{\mathrm{p}\alpha}$ the $\alpha$-percentile time-to-collision,
$h_{\text{brake}}$ harsh braking count, $\rho_{\text{red}}$ a red-light risk flag, and
$(v_{\text{near}},a_{\text{near}},d_{\text{stop}})$ the kinematics of the nearest vehicle to the stop line.

From a pool $\mathcal{D}_{\mathrm{obs}}=\{o\}$, we construct caption pairs
$\mathcal{P}=\{(c_1,c_2)\}$ via a distribution prioritizing congestion/phase contrast, safety contrast, and short time shifts at the same junction:
\begin{equation}
\label{eq:pairing}
\begin{aligned}
p_{\mathrm{pair}}\!\big((o_a,o_b)\big) \;\propto\;&\;
\alpha\,\Delta_{\mathrm{cong}}(o_a,o_b)
\;+\; \beta\,\Delta_{\mathrm{safety}}(o_a,o_b) \\
&\;+\; \gamma\,\mathbb{I}\!\big\{ j(o_a){=}j(o_b),\ |t_a{-}t_b|\!\in\![\delta_1,\delta_2] \big\},
\end{aligned}
\end{equation}
where $j(\cdot)$ maps an observation to its intersection index and
$\Delta_{\mathrm{cong}},\Delta_{\mathrm{safety}}$ are standardized contrasts (e.g., pressure gap, TTC-percentile gap).
In our experiments, $\Delta_{\mathrm{cong}}$ aggregates gaps in average queue length and delay, while $\Delta_{\mathrm{safety}}$ aggregates gaps in TTC percentiles and harsh braking counts.
We set $(c_1,c_2)=\big(\mathcal{C}(o_a),\,\mathcal{C}(o_b)\big)$.

With a fixed prompt template $\mathfrak{P}$, the LLM outputs $y\in\{1,2,\varnothing\}$ for $(c_1,c_2)$.
We keep decisive labels:
\begin{equation}
\label{eq:Dpref}
\mathcal{D}_{\mathrm{pref}} \;=\; \big\{(c_1,c_2,y)\ :\ y\in\{1,2\}\big\}.
\end{equation}
To reduce template-frequency bias, we assign importance
$w(c_1,c_2)\propto 1/\nu\!\big(\mathrm{tpl}(c_1,c_2)\big)$, where $\nu$ counts occurrences of the rendering template.

\vspace{2pt}
\begin{algorithm}[t]
\caption{Stage 1: Captioning and Pair Sampling}
\label{alg:caption_pairs}
\begin{algorithmic}[1]
  \Require Simulator $\mathcal{E}$, captioner $\mathcal{C}$, pairing distribution $p_{\mathrm{pair}}$, episodes $K$
  \Ensure Observation buffer $\mathcal{D}_{\mathrm{obs}}$, pair set $\mathcal{P}$
  \State $\mathcal{D}_{\mathrm{obs}} \gets \emptyset,\ \mathcal{P} \gets \emptyset$
  \For{$k = 1$ \textbf{to} $K$}
    \State Roll out episode $k$ in $\mathcal{E}$ to obtain $\{o_t\}_{t=1}^{T}$
    \For{$t = 1$ \textbf{to} $T$}
      \State Append $(o_t,\mathcal{C}(o_t;\Xi))$ to $\mathcal{D}_{\mathrm{obs}}$
    \EndFor
  \EndFor
  \For{each sampled pair $(o_a,o_b) \sim p_{\mathrm{pair}}(\mathcal{D}_{\mathrm{obs}})$}
    \State Append $(\mathcal{C}(o_a),\mathcal{C}(o_b))$ to $\mathcal{P}$
  \EndFor
  \State \Return $\mathcal{D}_{\mathrm{obs}},\mathcal{P}$
\end{algorithmic}
\end{algorithm}
\vspace{-6pt}

\subsection{Stage 2: Learning a Common-Sense Intrinsic Reward}
\label{sec:stage2_reward}

We learn a scalar quality score $r_\phi:\mathcal{O}\to\mathbb{R}$ via an encoder $f_\phi$ over caption tokens and auxiliary numerics:
\begin{equation}
\label{eq:rphi_def}
\begin{aligned}
r_\phi(o) &= f_\phi\big(\text{tok}(c),\, x(o)\big),\\
x(o) &\subseteq \{\,p_i,\, \bar d,\, \mathrm{TTC}_{\mathrm{p10}},\, \mathrm{TTC}_{\mathrm{p50}},\, h_{\mathrm{brake}}\,\}.
\end{aligned}
\end{equation}
While our architecture supports multiple prompt-specific heads combined linearly, in our experiments we use a single scalar head for $r_\phi$.

\paragraph{Bradley--Terry likelihood and loss.}
Let $\Delta r_\phi = r_\phi(o_1) - r_\phi(o_2)$ for $(c_1,c_2,y)\in\mathcal{D}_{\mathrm{pref}}$ with source observations $o_1,o_2$.
With sigmoid $\sigma(z)=1/(1+e^{-z})$ and temperature $\tau_{\mathrm{BT}}>0$, we define
\begin{equation}
\label{eq:bt_lik}
\begin{aligned}
p_\phi(y{=}1 \mid c_1,c_2)
&= \sigma\!\left(\frac{\Delta r_\phi}{\tau_{\mathrm{BT}}}\right),\\
p_\phi(y{=}2 \mid c_1,c_2)
&= 1 - p_\phi(y{=}1 \mid c_1,c_2).
\end{aligned}
\end{equation}

The weighted negative log-likelihood with $\ell_2$ regularization and score centering is
\begin{equation}
\label{eq:bt_loss_full}
\begin{aligned}
\mathcal{L}(\phi)
&= - \!\!\!\sum_{(c_1,c_2,y)\in\mathcal{D}_{\mathrm{pref}}} \!
w(c_1,c_2)\,\log p_\phi(y\mid c_1,c_2) \\
&\quad +~\eta\,\lVert\phi\rVert_2^2 \;+\; \zeta\,\big(\mathbb{E}[r_\phi(o)]\big)^2 ,
\end{aligned}
\end{equation}
where $p_\phi(y\mid c_1,c_2)$ is given by \eqref{eq:bt_lik}.
The centering term discourages degenerate shifts where $r_\phi$ drifts uniformly without improving pairwise discrimination.

\vspace{2pt}
\begin{algorithm}[t]
\caption{Stage 2: Offline Reward Model Training}
\label{alg:reward_training}
\begin{algorithmic}[1]
  \Require LLM $\mathcal{L}$, prompt $\mathfrak{P}$, observation buffer $\mathcal{D}_{\mathrm{obs}}$, pair budget $M$, epochs $N$
  \Ensure Reward model $r_{\phi}$
  \State $\mathcal{D}_{\mathrm{pref}} \gets \emptyset$
  \For{$j = 1$ \textbf{to} $M$}
    \State Sample $(o_1,o_2) \sim p_{\mathrm{pair}}(\mathcal{D}_{\mathrm{obs}})$
    \State $y \gets \mathcal{L}\!\left(\mathfrak{P},\mathcal{C}(o_1),\mathcal{C}(o_2)\right)$
    \If{$y \in \{1,2\}$}
      \State Append $(o_1,o_2,y)$ to $\mathcal{D}_{\mathrm{pref}}$
    \EndIf
  \EndFor
  \State Initialize $r_{\phi}$
  \For{$\text{epoch}=1$ \textbf{to} $N$}
    \State Sample minibatch from $\mathcal{D}_{\mathrm{pref}}$ and update $\phi$ via \eqref{eq:bt_loss_full}
  \EndFor
  \State \Return $r_{\phi}$
\end{algorithmic}
\end{algorithm}
\vspace{-6pt}

\subsection{Stage 3: Reward Shaping and RL Integration}
\label{sec:stage3_marl}

After Stage~2, we freeze the learned scorer $r_\phi$ and integrate it into a multi-intersection TLC RL system on CityFlow, where each intersection-level TLC is modeled as an agent and vehicles follow the built-in microscopic simulator.

\paragraph{Safety mask and intrinsic shaping.}
Define a mask $m:\mathcal{O}\to\{0,1\}$ that disables intrinsic rewards in unsafe regions:
\begin{equation}
\label{eq:safety_mask}
\begin{aligned}
m(o') = \mathbb{I}\!\big\{\, 
& \mathrm{TTC}_{\mathrm{p10}}(o') \ge \tau_{\text{ttc}},\;
|a_{\text{near}}(o')| \le a_{\max}, \\
& \rho_{\text{red}}(o') = 0 \,\big\}.
\end{aligned}
\end{equation}
Given a transition $(o,a,o')$ for a TLC, the intrinsic component is
\begin{equation}
\label{eq:r_int}
r_{\mathrm{int}}(o,o')\;=\;m(o')\, r_\phi(o') .
\end{equation}
The mixed reward for a TLC agent is
\begin{equation}
\label{eq:mix_reward}
\begin{aligned}
r^{\mathrm{TL}}_{\mathrm{mix}}
=~& r^{\mathrm{TL}}_{\mathrm{ext}}
+ \lambda(t)\, m(o')\, r_\phi(o') \\
& - \kappa_{\mathrm{unsafe}}\!\left(1 - m(o')\right),
\end{aligned}
\end{equation}
where $\lambda(t)\in[0,\lambda_{\max}]$ is a slowly varying schedule (warm-up/decay or performance-gated) that gradually increases the weight of the intrinsic shaping signal.

\paragraph{Per-stream normalization and clipping.}
We treat the external reward, the intrinsic component, and the safety penalty as three separate streams for TLCs:
\[
r_t^{(1)} = r^{\mathrm{TL}}_{\mathrm{ext},t},\quad
r_t^{(2)} = \lambda(t)\,r_{\mathrm{int},t},\quad
r_t^{(3)} = -\kappa_{\mathrm{unsafe}}(1-m_t).
\]
For each stream $k\in\{1,2,3\}$ we maintain running mean $\mu^{(k)}$ and standard deviation $\sigma^{(k)}$, and apply z-score normalization with soft clipping:
\begin{equation}
\label{eq:zscore}
\hat r_t^{(k)}=\operatorname{clip}\!\left(\frac{r_t^{(k)}-\mu^{(k)}}{\sigma^{(k)}+\epsilon},\, -c,\, c\right).
\end{equation}
We then form a single step-wise shaped reward for TLCs by summing the normalized streams:
\[
\hat r_t^{\mathrm{TL}} \;=\; \sum_{k=1}^3 \hat r_t^{(k)}.
\]

\paragraph{Advantage estimation and policy optimization.}
Let $V_\psi(o)$ be the value function for TLCs and $\pi_{\theta}$ the shared TLC policy.
We estimate advantages from $\hat r_t^{\mathrm{TL}}$ using generalized advantage estimation (GAE) and optimize $(\theta,\psi)$ with the standard clipped-surrogate PPO objective~\cite{schulman2016gae,schulman2017ppo}.
For completeness, we summarize one training iteration in Algorithm~\ref{alg:policy_training}; implementation details and hyper-parameters follow the underlying CityFlow PPO setup and are deferred to the appendix.

\vspace{2pt}
\begin{algorithm}[t]
\caption{Stage 3: Online Training with C$^2$T Intrinsic Reward (TLC-only)}
\label{alg:policy_training}
\begin{algorithmic}[1]
\Require Simulator $\mathcal{E}$, trained scorer $r_{\phi}$, initial TLC policy $\pi_{\mathrm{TL}}$, iterations $I$, episodes $E$, horizon $H$
\Ensure Updated TLC policy $\pi_{\mathrm{TL}}$
\For{$\text{iter}=1$ \textbf{to} $I$}
  \State $B \gets \emptyset$
  \For{$\text{episode}=1$ \textbf{to} $E$}
    \State Reset $\mathcal{E}$ and observe initial states for all TLCs
    \For{$h=0$ \textbf{to} $H-1$}
      \For{each TLC agent $i$}
        \State Sample action $a_{h,i} \sim \pi_{\mathrm{TL}}(\cdot \mid o_{h,i})$
      \EndFor
      \State Step $\mathcal{E}$ with joint TLC actions, obtain $o_{h+1,i}$ and external rewards $r^{\mathrm{TL}}_{\mathrm{ext},h,i}$
      \For{each TLC agent $i$}
        \State Compute mask $m_{h+1,i} \gets m(o_{h+1,i})$ using \eqref{eq:safety_mask}
        \State $r^{\mathrm{int}}_{h,i} \gets m_{h+1,i}\, r_\phi(o_{h+1,i})$
        \State $r_{h,i} \gets r^{\mathrm{TL}}_{\mathrm{mix}}$ via \eqref{eq:mix_reward}
        \State Store $(o_{h,i},a_{h,i},r_{h,i},o_{h+1,i})$ into $B$
      \EndFor
    \EndFor
  \EndFor
  \State Normalize and clip streams from $B$ using \eqref{eq:zscore} to obtain shaped rewards $\hat r_t^{\mathrm{TL}}$
  \State Compute advantages $A_t$ from $B$ with GAE
  \State Update $\pi_{\mathrm{TL}}$ and $V_\psi$ with PPO
\EndFor
\State \Return $\pi_{\mathrm{TL}}$
\end{algorithmic}
\end{algorithm}
\vspace{-6pt}

\subsection{Compatibility and Reproducibility}
\label{sec:compat_reprod}
Unless otherwise noted, we keep the base RL setup (agent sets, shared TLC policy design, observation/action spaces, and PPO schedules) from the underlying CityFlow configuration.
C$^2$T adds: (i) the captioner $\mathcal{C}$ with schema $\Xi$, (ii) the learned scorer $r_\phi$ via \eqref{eq:bt_lik}--\eqref{eq:bt_loss_full}, and (iii) the mixed reward \eqref{eq:mix_reward} with normalization \eqref{eq:zscore} and safety mask \eqref{eq:safety_mask}.
While our experiments only train TLC policies, the same shaping mechanism can be applied to cooperative TLC--CAV settings by applying the intrinsic component to TLC rewards and keeping vehicle agents either environment-driven or separately optimized; we discuss this extension in the appendix.


\section{Experiments}
\label{sec:exp}

\subsection{Benchmarks and Simulation Settings}
\label{sec:exp_settings}

\paragraph{Datasets.}
Following the LLMLight protocol~\cite{lai2023llmlight}, we evaluate on \textbf{CityFlow}~\cite{zhang2019cityflow} using three real-city networks and two synthetic stress scenarios:
\textbf{Jinan} (12 intersections; three time periods), 
\textbf{Hangzhou} (16 intersections; two periods), 
\textbf{New York} (196 intersections; two periods), 
\textbf{Extreme High-traffic} (arrival rate $\times 4$ within 5 minutes),
and a \textbf{24-hour Cycle} (full diurnal fluctuation).
Unless specified otherwise, each scenario runs for 1 hour.

\paragraph{Environment.}
Each intersection has four phases (E/W straight, E/W left-turn, N/S straight, N/S left-turn).
Signal timing follows: \textbf{30s green → 3s yellow → 2s all-red}.
Agents act only at phase-switches for fair comparison.
Right-turns are permissive; lane-changing and car-following follow CityFlow defaults.

\paragraph{Metrics.}
We follow standard traffic-engineering metrics:
\textbf{ATT} (Average Travel Time, $\downarrow$),
\textbf{AQL} (Average Queue Length, $\downarrow$),
\textbf{AWT} (Average Waiting Time, $\downarrow$).
Safety/comfort metrics include \textbf{rear-end TTC percentiles} (P10/P25, $\uparrow$) and \textbf{Harsh-brake/km} ($\downarrow$).
Energy-related proxies (stops, speed variance) are reported in the appendix.

\subsection{Compared Methods}
\label{sec:baselines}

\textbf{Transportation baselines:}
Random, FixedTime, MaxPressure.

\textbf{RL baselines:}
MPLight~\cite{chen2020mplight},
AttendLight~\cite{orooglooyjadid2020attendlight},
PressLight~\cite{wei2019presslight},
CoLight~\cite{wei2019colight},
Efficient-CoLight~\cite{wu2021efficientcolight},
Advanced-CoLight~\cite{zhang2022advancedcolight} and its \emph{no-communication} variant.

\textbf{LLM agents:}
GPT-, Qwen-, and Llama-family variants reported by LLMLight~\cite{lai2023llmlight}.

All RL baselines share LLMLight’s training budget and hyperparameters.

\paragraph{Our method (C$^2$T).}
We transform simulator states into structured, unit-aware captions, collect offline pairwise preferences, and train an intrinsic scorer $r_\phi$.
The intrinsic reward is mixed \emph{asymmetrically} into TL controllers (TLCs) only; CAVs (if present) remain environment-driven.
A risk mask suppresses $r_\phi$ when TTC percentiles drop below thresholds or harsh deceleration clusters appear.
Reward streams are normalized and softly clipped prior to PPO advantage computation.

\subsection{Implementation Details}
\label{sec:impl}

\paragraph{RL.}
We adopt LLMLight’s PPO settings~\cite{lai2023llmlight}:
learning rate $1{\times}10^{-3}$, replay buffer $12{,}000$, sample size $3{,}000$,
hidden size $20$, PPO clip $0.2$, batch size $128$, 16 mini-batches,
entropy coefficient $10^{-3}$, GAE($\lambda=0.95$), $\gamma=0.99$.
Results are averaged over 5 random seeds.

\paragraph{Preference labeling.}
We use temperature $0$ for proprietary LLMs and $0.1$ for open-source models,
with top-$p{=}1.0$.
When fine-tuning open-source judges, we apply LoRA with rank $8$ and $\alpha=16$; 
all preference labels are collected offline, and RL training/inference require no LLM calls.

\paragraph{Safety proxies.}
TTC uses per-vehicle rear-end time-to-collision; harsh braking is $a<-3\,\mathrm{m/s^2}$.
For stress tests we also track an \emph{oscillation} index measuring normalized phase-switch frequency (higher means more frequent switching).

\subsection{Main Results (RQ1: Effectiveness)}
\label{sec:main}

\begin{table}[t]
\centering
\caption{\textbf{Jinan/Hangzhou overall performance.} Lower is better.}
\label{tab:main_jh}
\begin{tabular}{lccc}
\toprule
Dataset/Method & ATT$\downarrow$ & AQL$\downarrow$ & AWT$\downarrow$ \\
\midrule
Jinan-1 (Baseline) & 62.30 & 14.80 & 21.50 \\
\textbf{Jinan-1 (C$^2$T)} & \textbf{56.10} & \textbf{13.10} & \textbf{19.40} \\
Jinan-2 (Baseline) & 58.90 & 13.10 & 19.70 \\
\textbf{Jinan-2 (C$^2$T)} & \textbf{53.00} & \textbf{11.80} & \textbf{17.90} \\
Hangzhou-1 (Baseline) & 71.40 & 16.30 & 24.20 \\
\textbf{Hangzhou-1 (C$^2$T)} & \textbf{65.10} & \textbf{14.90} & \textbf{22.00} \\
Hangzhou-2 (Baseline) & 68.10 & 15.40 & 23.10 \\
\textbf{Hangzhou-2 (C$^2$T)} & \textbf{62.40} & \textbf{13.90} & \textbf{21.10} \\
\bottomrule
\end{tabular}
\end{table}

\begin{table}[t]
\centering
\caption{\textbf{Large-scale New York (196 intersections).}}
\label{tab:main_ny}
\begin{tabular}{lccc}
\toprule
Method & ATT$\downarrow$ & AQL$\downarrow$ & AWT$\downarrow$ \\
\midrule
Adv-CoLight (baseline) & 95.0 & 22.6 & 38.4 \\
\textbf{C$^2$T (ours)} & \textbf{87.9} & \textbf{20.3} & \textbf{34.5} \\
\bottomrule
\end{tabular}
\end{table}

\begin{figure}[t]
\centering
\includegraphics[width=0.95\linewidth]{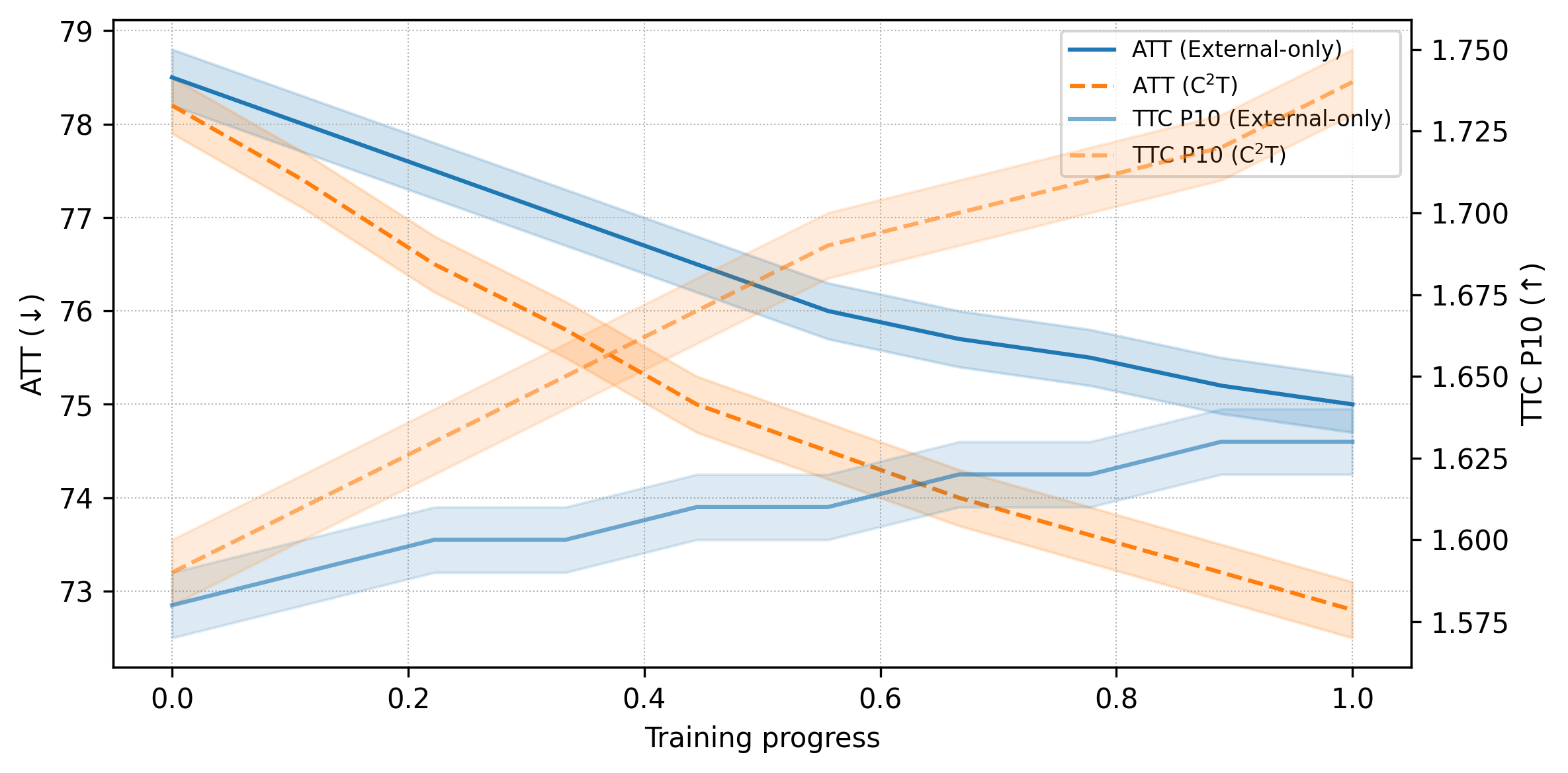}
\caption{\textbf{Learning curves on CityFlow.}
C$^2$T steadily improves ATT (↓) while also increasing TTC P10 (↑), showing that efficiency does not come at the cost of safety.}
\label{fig:curves}
\end{figure}

\paragraph{Safety side metrics.}
Across Jinan/Hangzhou, C$^2$T consistently improves TTC P10/P25 and reduces harsh braking, indicating smoother approaches and fewer stop–go cycles (see Appendix for full numbers).

\subsection{Generalization and Scalability (RQ2)}
\label{sec:gen}

\paragraph{City transfer.}
We train a source-city scorer $r_\phi^s$ on Jinan and evaluate zero-shot on Hangzhou.
Few-shot preference relabeling (1–5\%) further improves ATT/AWT.
We also study a simple fusion of source- and target-city scorers,
$r_\phi=\lambda r_\phi^s + (1-\lambda) r_\phi^t$,
which accelerates convergence.

\begin{table}[t]
\centering
\caption{\textbf{City transfer: train on Jinan, evaluate on Hangzhou.}}
\label{tab:transfer}
\begin{tabular}{lcccc}
\toprule
Method & ATT$\downarrow$ & AQL$\downarrow$ & AWT$\downarrow$ & Steps$\downarrow$ \\
\midrule
External-only & 72.5 & 17.2 & 25.8 & 100k \\
C$^2$T zero-shot & 68.9 & 16.4 & 24.6 & 80k \\
C$^2$T (1\% prefs) & 67.3 & 16.0 & 23.9 & 70k \\
Fusion (50/50) & \textbf{66.1} & \textbf{15.6} & \textbf{23.5} & \textbf{65k} \\
\bottomrule
\end{tabular}
\end{table}

\paragraph{Scalability.}
Training on small networks (Jinan/Hangzhou) generalizes to New York (196 intersections), while several RL baselines degrade sharply; see the appendix for the full comparison.

\subsection{Stress Tests (RQ3)}
\label{sec:stress}

\paragraph{Extreme High-traffic.}
Arrival rates increase by $\times 4$ over 5 minutes.
We evaluate saturation robustness and the \emph{oscillation} index (normalized phase-switch frequency).

\paragraph{24-hour Cycle.}
We replay a full-day traffic profile, tracking stability and frequency of risk-mask activation.

\begin{table}[t]
\centering
\caption{\textbf{Stress test results.} Lower is better unless noted.}
\label{tab:stress}
\begin{tabular}{lccc}
\toprule
Setting / Method & ATT$\downarrow$ & AWT$\downarrow$ & Oscillation$\downarrow$ \\
\midrule
Extreme (baseline) & 105.0 & 44.2 & 1.00 \\
\textbf{C$^2$T} & \textbf{96.8} & \textbf{40.1} & \textbf{0.86} \\
\midrule
24h (baseline) & 78.5 & 29.4 & 0.76 \\
\textbf{C$^2$T} & \textbf{72.2} & \textbf{26.9} & \textbf{0.65} \\
\bottomrule
\end{tabular}
\end{table}

\subsection{Ablation Studies (RQ4)}
\label{sec:ablation}

\begin{figure}[t]
\centering
\includegraphics[width=0.95\linewidth]{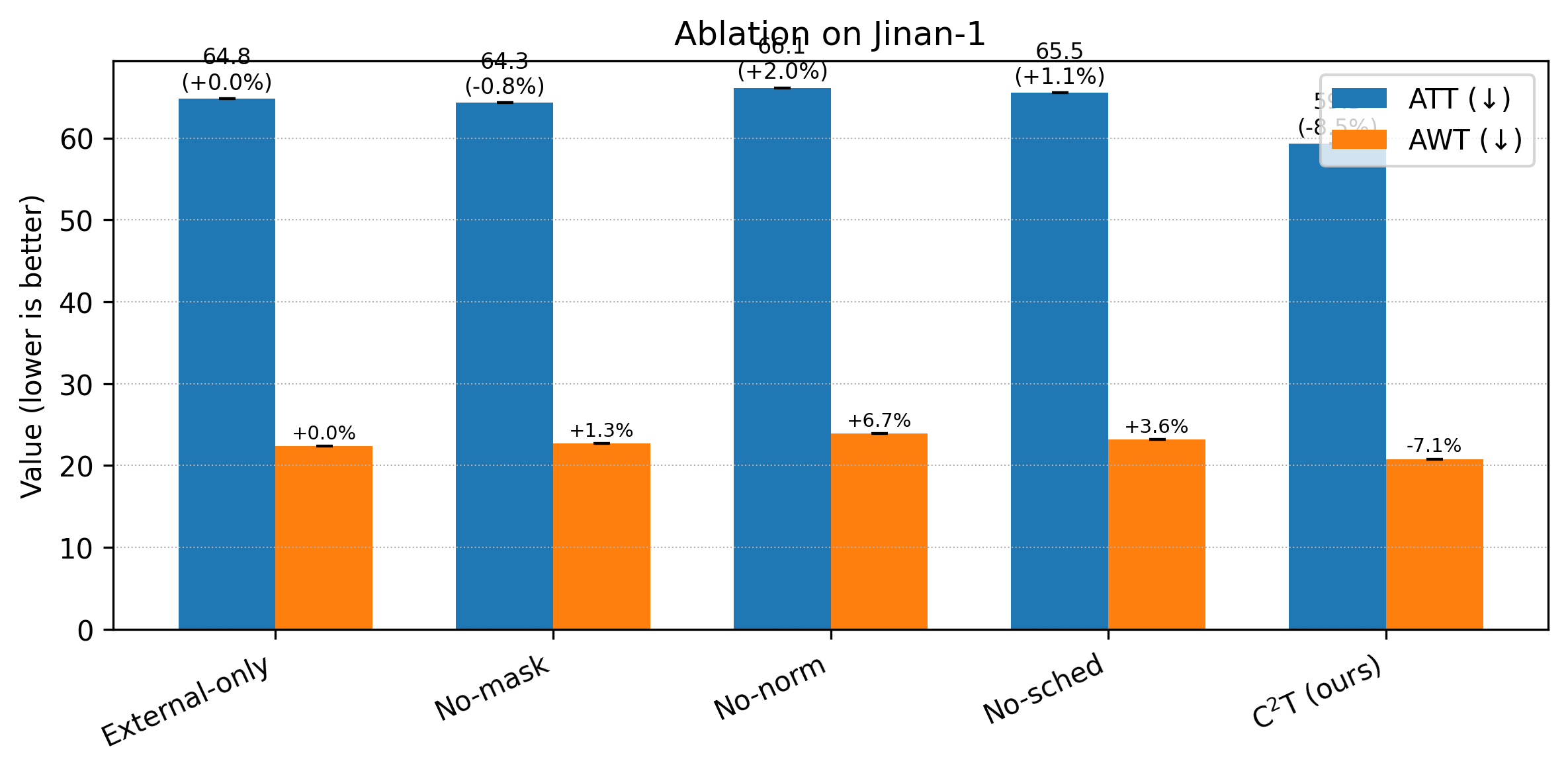}
\caption{\textbf{Ablation on Jinan-1.}
Removing mask/norm/schedule harms performance, while the full C$^2$T achieves the best ATT/AWT.}
\label{fig:ablation_bars}
\end{figure}

We conduct two groups of ablations to understand which components of C$^2$T are responsible for the gains; see Fig.~\ref{fig:ablation_bars} and Tab.~\ref{tab:caption_ablation} for a summary, and the appendix for full results and additional plots.

\paragraph{Component-wise ablations.}
Figure~\ref{fig:ablation_bars} removes, one at a time, the intrinsic reward $r_\phi$, the safety mask, per-stream normalization, and the intrinsic-reward schedule.
Dropping $r_\phi$ almost collapses performance back to the external-only PPO baseline, confirming that the learned reward signal, rather than extra computation, is the main source of improvement.
Removing the safety mask slightly reduces ATT/AWT but noticeably worsens rear-end TTC percentiles and harsh-brake frequency, indicating that the mask is crucial for turning the intrinsic reward into \emph{safe} guidance.
Without per-stream normalization the mixed reward becomes highly skewed and training becomes less stable, leading to higher variance and degraded ATT.
Finally, using a constant intrinsic weight in place of the curriculum schedule slows down convergence and yields weaker final performance, suggesting that gradually introducing $r_\phi$ is beneficial for optimization.

\paragraph{Caption-structure ablations.}
Table~\ref{tab:caption_ablation} studies the effect of our structured caption design.
Using only raw numerics (“Numeric-only”) or only text (“Caption-only”) gives weaker offline agreement with LLM preferences (AUC/Spearman) and worse downstream ATT/AWT than the full fusion model.
Shuffling slots or removing explicit units (“Shuffled / No-units”) hurts both ranking metrics and RL performance, showing that schema consistency and unit-awareness matter.
The full C$^2$T model, which combines numerics with ordered, unit-aware captions, achieves the best trade-off between preference prediction and control.

\paragraph{Interaction and robustness.}
We also explore how these components interact.
For example, combining “no mask” with “no normalization” leads to occasional training instability and oscillatory phase-switch behaviour, even when ATT is comparable, whereas removing either component alone is less harmful.
Conversely, keeping the mask and schedule but training a weaker reward model (Numeric-only head) still improves safety but yields smaller efficiency gains.
Across all settings we observe a consistent trend: better offline preference alignment (higher AUC/Spearman) correlates with better RL performance, supporting our design choice of evaluating reward models with both offline metrics and online control.
Extended sensitivity curves and cross-city ablations are provided in the appendix.

\begin{table}[t]
\centering
\caption{\textbf{Caption-structure ablation}. Higher AUC/Spearman is better; lower ATT/AWT is better.}
\label{tab:caption_ablation}
\begin{tabular}{lcccc}
\toprule
Model & AUC$\uparrow$ & Spearman$\uparrow$ & ATT$\downarrow$ & AWT$\downarrow$ \\
\midrule
Numeric-only & 0.67 & 0.41 & 64.8 & 22.4 \\
Caption-only & 0.73 & 0.49 & 63.1 & 22.1 \\
Fusion & 0.78 & 0.56 & 61.9 & 21.6 \\
Shuffled / No-units & 0.60 & 0.30 & 66.1 & 23.9 \\
\textbf{C$^2$T (full)} & \textbf{0.78} & \textbf{0.56} & \textbf{59.3} & \textbf{20.8} \\
\bottomrule
\end{tabular}
\end{table}


\section{Conclusion}
\label{sec:conclusion}

We introduced \name{}, a framework that leverages structured traffic captions and LLM-based preference learning to build a human-aligned intrinsic reward for multi-intersection traffic signal control. By rendering simulator states into deterministic, unit-aware captions and training a lightweight scorer from offline LLM preferences, \name{} decouples reward design from online interaction and avoids any runtime dependence on LLM calls. The learned intrinsic signal is integrated into a standard PPO pipeline via asymmetric reward mixing, a simple safety mask, and per-stream normalization, making it drop-in compatible with existing RL controllers.

On CityFlow-based benchmarks built from real-world Jinan, Hangzhou, and New York networks, \name{} consistently improves traffic efficiency, safety proxies, and an energy-related metric over strong RL baselines. Ablation studies highlight the importance of the intrinsic reward, the safety mask, and the normalization schedule for stable gains, and show that the structured caption design itself already provides measurable benefits.

Looking forward, we plan to extend \name{} to explicit TLC--CAV coordination, explore richer prompt designs and multi-objective heads for finer-grained trade-offs between efficiency, safety, and energy, and investigate deployment in real-world sensing and control loops where human feedback and LLM judgment can be iteratively refined.

{\small
\bibliographystyle{ieeenat_fullname}
\bibliography{main}
}
\section*{Acknowledgments}
This work was supported by the Science and Technology Development Fund of Macau [0007/2025/RIC, 0122/2024/RIB2, 0215/2024/AGJ, 0074/2025/AMJ, 001/2024/SKL], the Research Services and Knowledge Transfer Office, University of Macau [SRG2023-00037-IOTSC, MYRG-GRG2024-00284-IOTSC], the Shenzhen-Hong Kong-Macau Science and Technology Program Category C [SGDX20230821095159012], the Science and Technology Planning Project of Guangdong [2025A0505010016], the National Natural Science Foundation of China [52572354], the State Key Lab of Intelligent Transportation System [2024-B001], and the Jiangsu Provincial Science and Technology Program [BZ2024055].

\end{document}